\documentclass{article}

\usepackage{PRIMEarxiv}

\usepackage[utf8]{inputenc} 
\usepackage[T1]{fontenc}    
\usepackage{hyperref}       
\usepackage{url}            
\usepackage{booktabs}       
\usepackage{amsfonts}       
\usepackage{nicefrac}       
\usepackage{microtype}      
\usepackage{lipsum}
\usepackage{graphicx}
\graphicspath{{media/}}     
\usepackage{booktabs}
\usepackage{multirow}
\usepackage{adjustbox}
\usepackage{threeparttable}
\usepackage{pdflscape}
\usepackage[square,sort]{natbib}
  
\title{Automatically Detecting Confusion and Conflict During Collaborative Learning Using Linguistic, Prosodic, and Facial Cues
\thanks{\textit{\underline{Please note}}: 
\textbf{This is a manuscript submitted the journal of User Modeling and User-Adapted Interaction. Year of 2024. Under-review.}} 
}

\author{
  Yingbo Ma \\
  University of Florida \\
  Gainesville, FL, USA \\
  \texttt{yingbo.ma@ufl.edu} \\
  \And
  Yukyeong Song \\
  University of Florida \\
  Gainesville, FL, USA \\
  \texttt{y.song1@ufl.edu} \\
  \And
  Mehmet Celepkolu \\
  University of Florida \\
  Gainesville, FL, USA \\
  \texttt{mckolu@ufl.edu} \\
  \And
  Kristy Elizabeth Boyer \\
  University of Florida \\
  Gainesville, FL, USA \\
  \texttt{keboyer@ufl.edu} \\
  \And
  Eric Wiebe \\
  North Carolina State University \\
  Raleigh, NC, USA \\
  \texttt{wiebe@ncsu.edu} \\
  \And
  Collin F. Lynch \\
  North Carolina State University \\
  Raleigh, NC, USA \\
  \texttt{cflynch@ncsu.edu} \\
  \And
  Maya Israel \\
  University of Florida \\
  Gainesville, FL, USA \\
  \texttt{misrael@coe.ufl.edu}
}

\begin{document}
\maketitle

\begin{abstract}
During collaborative learning, confusion and conflict emerge naturally. However, persistent confusion or conflict have the potential to generate frustration and significantly impede learners' performance. Early automatic detection of confusion and conflict would allow us to support early interventions which can in turn improve students' experience with and outcomes from collaborative learning. Despite the extensive studies modeling confusion during \textit{solo} learning, there is a need for further work in \textit{collaborative} learning. This paper presents a multimodal machine-learning framework that automatically detects confusion and conflict during collaborative learning.  We used data from 38 elementary school learners who collaborated on a series of programming tasks in classrooms. We trained deep multimodal learning models to detect confusion and conflict using features that were automatically extracted from learners' collaborative dialogues, including (1) language-derived features including TF*IDF, lexical semantics, and sentiment, (2) audio-derived features including acoustic-prosodic features, and (3) video-derived features including eye gaze, head pose, and facial expressions.  Our results show that multimodal models that combine semantics, pitch, and facial expressions detected confusion and conflict with the highest accuracy, outperforming all unimodal models. We also found that prosodic cues are more predictive of conflict, and facial cues are more predictive of confusion. This study contributes to the automated modeling of collaborative learning processes and the development of real-time adaptive support to enhance learners’ collaborative learning experience in classroom contexts.
\end{abstract}


\section{Introduction}
Collaborative learning refers to the coordinated attempt between two or more learners to solve a problem, complete a task, or create a shared product \citep{lai2011collaboration}. While collaborating, learners bring different ideas, defending and evaluating their perspectives \citep{pena2004analyzing} to produce a shared solution \citep{van2006social}. Effective collaboration has a number of benefits for learners, such as promoting critical thinking \citep{lee2015facilitating,loes2017collaborative}, increasing self-esteem \citep{sultan2012comparison,adene2021effectiveness}, and developing social skills \citep{tolmie2010social,laal2012benefits}. 

However, simply placing learners in groups and assigning them a task does not guarantee that learners will collaborate effectively \citep{soller2001supporting}. In order to derive the most benefit from collaborative learning, the team members should actively engage in cognitive and social processes for knowledge co-construction \citep{sangin2011facilitating,tran2013theoretical,vuopala2019knowledge}. During this dynamic interaction, confusion and conflict emerge naturally during collaboration and may pose particular challenges for learners \citep{kwon2014group}. Confusion arises when learners find mismatches between new information and their prior knowledge \citep{lehman2012confusion}, and conflict refers to disagreements among team members \citep{ayoko2008influence}. Although prior research has shown that confusion and conflict may be beneficial for learning \citep{d2014confusion,lee2015collaboration}, persistent confusion can generate frustration or boredom \citep{graesser2012emotions,lodge2018understanding}. Similarly, conflict can generate stress, decreasing satisfaction and mutual trust among group members \citep{naykki2014socio,lee2015collaboration,tsan2021collaborative}.

One solution to this challenge is to build adaptive intelligent systems to support collaborative learning and mitigate these negative outcomes \citep{laurillard2009pedagogical,magnisalis2011adaptive}. Such systems provide tailored scaffolding to learners by analyzing the dynamics of their group interaction in real-time \citep{moreno2013teaching,walker2014adaptive}. By automatically detecting conflict and confusion  during collaboration, intelligent systems can offer timely assistance to learners to resolve confusion and conflict. An important precursor to this goal is the automatic modeling of confusion and conflict during collaborative learning.

While previous studies have developed automated techniques and models to detect learners' confusion in individual learning contexts,  collaborative contexts are underexplored. \citet{d2008automatic} developed models that detected individual learners' affective states (e.g., confusion, flow, boredom) using dialogue features extracted from their conversations with an intelligent tutoring system; \citet{grawemeyer2017affective} developed an intelligent tutoring system that provided formative feedback to learners adapted to their affective states based on their speech and learning progress. However, there remains a gap in modeling confusion in collaborative learning scenarios. Though there has been work on detecting conflict in collaboration based on linguistic analysis, such as by detecting the presence of specific words \citep{prata2009detecting} and estimating sentiment from learners' chat messages \citep{lescano2021detecting}, there is no prior research specifically focused on automatically detecting conflict using other modalities of information from learners' interactions, such as speech and facial expressions. On the other hand, prior research on multimodal modeling of collaborative learning has analyzed learners' interactions across multiple modalities, including speech, facial expressions, body gestures \citep{subburaj2020multimodal,schneider2021gesture}. This technique has shown superior performance in attempts to understand students’ learning \citep{andrade2017understanding}, predict learning performance \citep{spikol2018supervised}, and construct models of students’ interactions \citep{sharma2020predicting}.

Building on these previous efforts, our goal is to investigate the multimodal modeling of confusion and conflict by analyzing learners' collaborative dialogues, including speech, eye gaze, and facial expressions. We specifically address two research questions:

\begin{itemize}
    \item \textbf{RQ 1}: What unimodal features are predictive of confusion and conflict during learners' collaborative interactions?
    \item \textbf{RQ 2}: Does multimodal feature fusion help improve detection performance compared to using unimodal features? If so, what are the best multimodal combinations among those considered?
\end{itemize}
To answer these research questions, we analyzed audio and video data collected from 38 elementary school learners who worked in pairs on collaborative coding activities. We extracted multimodal features from learners' collaborative dialogues, including 1)language-derived features including term frequency * inverse document frequency (TF*IDF) \citep{kusner2015word}, lexical semantics generated by pre-trained RoBERTa \citep{liu2019roberta}, and sentiment \citep{vivian2016method}, (2) audio-derived features including acoustic-prosodic features such as energy and pitch extracted with openSMILE \citep{eyben2010opensmile}, and (3) video-derived features including eye gaze, head pose, and facial expressions extracted with OpenFace \citep{baltrusaitis2018openface}.

To answer RQ1, we trained machine learning models on detecting confusion and conflict using each of those unimodal features extracted from learners' collaborative dialogues. The experimental results showed that semantics from language modality are predictive of both confusion and conflict; meanwhile, pitch from prosodic features from audio modality is more predictive of conflict, and facial expressions from video modality are more predictive of confusion. To answer RQ2, we evaluated models trained with different sets of multimodal features. The experimental results showed that multimodal models that combine lexical semantics, pitch, and facial expressions detected confusion and conflict with the highest accuracy, outperforming all unimodal models.

This study makes two main contributions to the domain: (1) we present the results from a sequence of experiments evaluating both a variety of predictive features and a selection of machine learning and deep sequential learning models; and (2) we incorporate multimodal features to achieve state-of-the-art accuracy in detecting confusion and conflict during collaborative learning. Our experimental results can serve as strong baselines for future research on the automatic modeling of collaborative learning processes and contribute to the development of real-time adaptive support to enhance the collaborative learning experience in classroom contexts.

The rest of the paper is organized as follows. Section~\ref{sec2} presents the background for this study; Section~\ref{sec3} describes the dataset used in this study; Section~\ref{sec4} details the features we investigated; Sections~\ref{sec5},~\ref{sec6} and~\ref{sec7} elaborate on the peer satisfaction prediction models; Section~\ref{sec6} presents the experimental settings and results; Section~\ref{sec8} discusses the implications of experimental results; and finally, Section~\ref{sec9} concludes the paper and discusses future work.

\section{Background}\label{sec2}
In framing this work, we drew on two key bodies of work. First, we briefly outline the theoretical foundations of how confusion and conflict impact learning, which highlight the importance of modeling these constructs during collaborative learning. Second, we describe the prior work on modeling confusion and conflict during collaborative learning, which lays the groundwork for the multimodal machine learning framework we used in this study for automatically detecting confusion and conflict.

\subsection{Confusion and Conflict during Learning}
Learners may experience a variety of cognitive-affective states when they are assigned difficult tasks to solve, including confusion, frustration, boredom, engagement/flow, curiosity, anxiety, delight, and surprise \citep{graesser2012emotions}. Among these affective states, confusion is particularly interesting because it is often correlated with learning gains \citep{craig2004affect,lehman2012confusion}. \cite{d2012dynamics} proposed a model of affective dynamics during learning that illustrates how those states evolve, morph, interact, and influence learning. According to the model, confusion is a key signature of cognitive disequilibrium \citep{vanlehn2003only} that arises when learners find mismatches between new information and their prior knowledge. Confusion may be beneficial for learning because learners are often motivated to actively engage in resolving the confusion and restoring equilibrium, which leads to a deeper understanding of the problem \citep{d2014confusion}. However, if confusion persists (e.g., learners are unable to resolve the confusion), it has the potential to result in frustration, leading learners to feel bored and disengaged \citep{baker2010better,bosch2017affective}. By limiting the persistence of confusion, educational interventions that support learners in resolving confusion hold the potential to enhance learning outcomes \citep{sabourin2013affect,di2020confused}.

Similar to confusion, conflict in collaborative learning contexts may also result in beneficial outcomes for learning, provided that learners receive appropriate support for conflict resolution \citep{lai2011collaboration,abbasi2017conflict}. Conflict in this work refers to disagreements among collaborative learners \citep{ayoko2008influence,kwon2014group} and can be categorized into two types: task conflict and relationship conflict \citep{clark2003collaborative}. Task conflict refers to disagreements between learners regarding opinions, perspectives, and ways of processing the task information \citep{garcia2007intergroup}. This type of conflict may be beneficial for learning because learners may benefit from being challenged by their partners, listening to each other’s arguments, and building upon each other’s ideas \citep{lee2015collaboration,zakaria2021identifying}. It has the potential to promote cohesion among team members \citep{jiang2013emotion} and increase group performance \citep{lee2015collaboration}. Relationship conflict, on the other hand, refers to disagreements between learners caused by issues unrelated to the task, and is often manifested as social friction and personality clashes \citep{de2003task}. This type of conflict can generate stress and negative emotions among learners, which can significantly impede learners’ group performance \citep{naykki2014socio,tsan2021collaborative}. Regardless of the type of conflict, it is important to provide learners with appropriate scaffolds in order to help them resolve both task and relationship related conflicts and improve their collaborative learning experience \citep{li2013cooperative,brown2018guided}.

This substantial body of theoretical evidence and empirical data point to the ways in which both confusion and conflict may benefit collaborative learners by promoting deeper understanding if they are appropriately regulated and resolved. However, this work also identifies how persistent confusion and conflict can generate frustration and significantly impede learners’ group performance. The present study draws on these theoretical foundations, and it advances the research by automatically modeling conflict and confusion in collaborative settings.

\subsection{Automatic Modeling of Confusion and Conflict}
Studies in affective computing often focus on multimodal approaches that combine multiple modalities such as facial expressions, vocal cues, body language, and physiological signals \citep{beyan2021modeling,li2021eeg,kalatzis2022emotions}. In recent years, the field of affective computing has been rapidly growing, with new methods and techniques continuing to emerge for emotion recognition \citep{wang2022systematic}. Deep learning methods, such as convolutional or recurrent neural network (CNN or RNN)-based feature representation techniques, have demonstrated early success in the field \citep{zhao2021multimodal}. Apart from these, Transformers \citep{vaswani2017attention}, which were originally proposed for which was originally proposed for machine translation in natural language processing (NLP), have continued to surpass the performance of the existing state-of-the-art deep sequential models. \cite{ma2021facial} utilized Visual Transformers with an attentional feature fusion strategy that aggregates both global and local relationships among multimodal features.

In the last two decades, modeling learners' affective states, including confusion, has received great interest among researchers and has been regarded as one of the most important aspects of learner modeling \citep{calvo2010affect,desmarais2012review,mejbri2022trends}. Extensive work has been dedicated to modeling individual learners' confusion when they interact with intelligent tutoring systems. For example, \citet{d2007toward} designed and developed Autotutor, an intelligent tutoring system sensitive to a learners' affective state. The system detects learners' affect (e.g., boredom, confusion, frustration) by using their reaction time and specific words and phrases (e.g., ``I don't know''). \citet{frasson2010managing} proposed an automatic approach to detect individuals' emotions while interacting with an intelligent tutoring system by measuring their brain activity using physiological data, such as electroencephalograms (EEG). \citet{wiggins2015javatutor} used Javatutor, an intelligent tutoring system for introductory computer science, which worked alongside students to support them through both cognitive (skills and knowledge) and affective (emotion-based) feedback. Similar to these intelligent systems, researchers have developed affect-aware systems integrated into game-based learning environments \citep{sabourin2011generalizing,sabourin2013affect,bosch2015automatic,plass2015foundations,hamari2016challenging}, or into other online learning platforms \citep{dillon2016student,xing2019beyond,moreno2020re,liu2022automated}. 

Confusion can be detected through multiple communicative modalities such as speech, facial expressions, and gestures. For example, a large body of research has related confusion to specific types of motion in learners' facial muscles, including Brow Lowering (AU4), Eyelid Tightening (AU7), and Lip Tightener (AU23) \citep{d2009multimethod,grafsgaard2011predicting,grafsgaard2013automatically,padron2016identification,ma2022detecting}.
In a similar vein, \citet{arroyo2009emotion} successfully detected learners' affective states, including confusion, through a series of physiological sensors (e.g., pressure mouse, posture analysis seat) integrated with intelligent tutors in a classroom setting. Confusion has also been connected to indicative conversational cues, such as prolonged response time and signal words or phrases \citep{d2007toward,d2008automatic}. Also, there have been research efforts towards the multimodal modeling of confusion, as well as other affective states during learning, by combining both facial and body movements with gesture cues \citep{shan2007beyond,karg2013body}. In alignment with this prior work, the present study leverages the advantage of multimodal learning analytics to automatically detect learners' confusion moments during collaborative learning by using their linguistic, acoustic, and visual features. 

Prior work on modeling conflict during collaborative learning has mainly focused on linguistic analysis. For example, \cite{lescano2021detecting} analyzed a dataset of chat messages from 74 student groups collaborating on their homework assignments. The authors used the sentiment features estimated from students' chat messages and then built machine learning algorithms to classify these messages as either conflict or non-conflict. They found that their method achieved high accuracy in detecting conflict in chat messages. \cite{vandenberg2021prompting} used Epistemic Network Analysis (ENA) to analyze the dialogues of 62 fourth-grade learners while engaging in collaborative coding tasks. The students were given interventions that aimed to prompt better collaboration during their programming tasks. The results indicated that the interventions fostered better collaborative behaviors among students, in which they asked questions, negotiated conflicts, and offered alternative ideas, than a control condition. In another study, \cite{vandenberg2022remember} used ENA to analyze elementary students' collaborative and regulated discourse during collaborative programming tasks. The authors found that students' self-efficacy and CS conceptual understanding affected their discourse. The results suggested that upper elementary students need lack conflict resolution skills and they need support to learn about productive disagreement and how to peer model.

In summary, confusion, as an important affective state, has been extensively studied during \textit{sole} learning; conflict has been mainly studied through single modality work in prior work. Very little has been done to investigate the automatic modeling of learners' confusion and conflict in \textit{collaborative} learning environments using multi-modalities of information from learners’ interactions. In alignment with these prior works, the present study leverages the advantage of multimodal learning analytics to automatically detect learners' confusion moments during collaborative learning by using their linguistic, acoustic, and visual features. 

The work reported in this paper is novel in several aspects compared to previous studies: (1) very little work has modeled learners' confusion and conflict in collaborative learning settings with multimodal features. In contrast, our work leverage multimodal features from lexical, audio, and video modalities to detect confusion and conflict during collaboration; we also contrast standard classifiers with deep sequential learning approaches. (2) Previous studies used multimodal approaches (e.g., \citep{chango2022review}) to analyze collaborative learning with traditional feature fusion methods. To the best of our knowledge, the work reported here is the first to investigate the performance of multimodal modeling with recent state-of-the-art neural network based feature fusion methods. These methods have proven to be more effective than traditional feature fusion methods on a series of learning tasks such as sentiment analysis \citep{zadeh2017tensor} and image classification \citep{mohla2020fusatnet}.

\section{Dataset}\label{sec3}
\subsection{Participants and Collaborative Setting}
Our dataset was collected from 38 learners in 4th grade classrooms in a rural elementary school in the Summer of 2022 in the southeastern United States. Out of the 38 learners, 21 were female, 12 were male, and 5 preferred not to report their gender. The distribution of race/ethnicity of these learners included 72\% self-reporting as White/Caucasian, 15\% Hispanic/Latinx, 9\% Black/African American, 4\% Multiracial, and 1\% Other. The mean age was 10.1 with ages ranging from 8 to 12.

\begin{figure*}[!tbp]
    \centering
    \includegraphics[width=\textwidth]{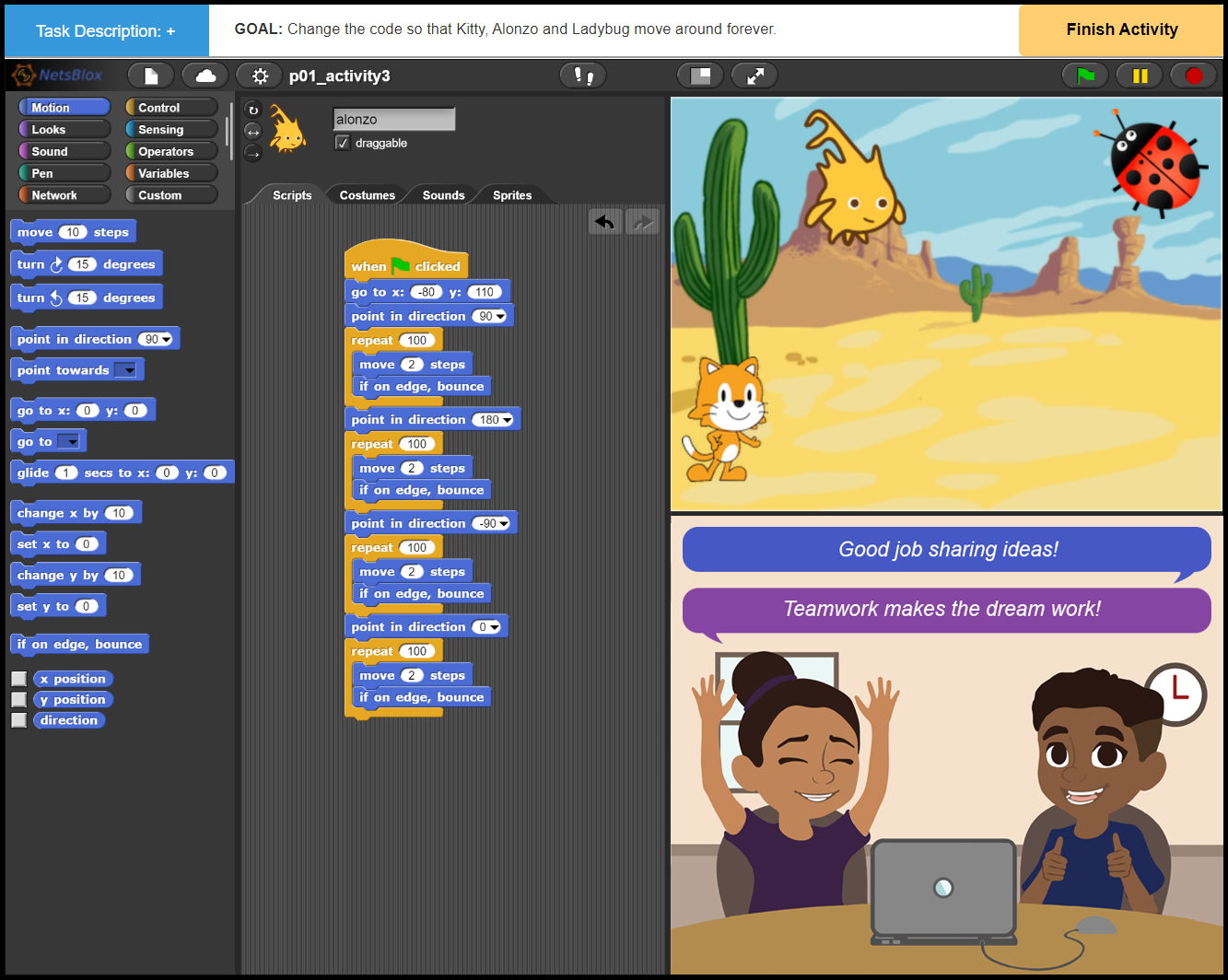}
    \caption{The FLECKS learning environment, which embeds virtual learning companions (lower right) in the block-based coding environment.}
    \label{fig:flecks}
\end{figure*}

The learners followed the pair programming paradigm, a popular framework for collaborative learning in computer science \citep{williams2002support}. In pair programming, the \textit{driver} is responsible for writing the code and implementing the solution, while the \textit{navigator} provides support by catching mistakes and providing feedback on the in-progress solution \citep{celepkolu2018importance}. During the pair programming activities of our study, each pair of learners shared one computer and switched roles between the driver and the navigator during the science-simulation coding activity using the FLECKS block-based coding environment, which extends Netsblox \citep{broll2017visual}. FLECKS (See Figure~\ref{fig:flecks}) includes two pedagogical virtual agents designed to foster collaborative learning experiences by modeling good collaboration practices for children and helping them learn about resolving potential conflicts with their partners. 

\begin{figure*}[!tbp]
    \centering
    \includegraphics[width=.6\textwidth]{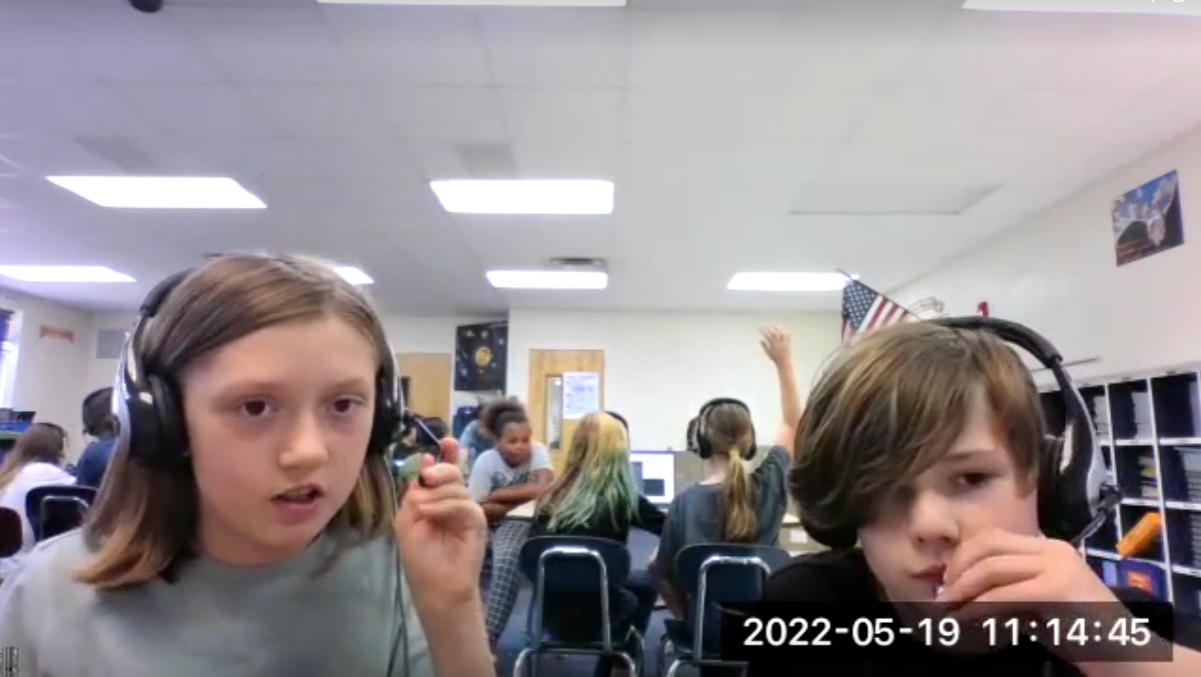}
    \caption{Two elementary school learners collaborating on a pair programming task. In the captured moment, the learner in the left side of the frame is the \textit{navigator} and the learner on the right is the \textit{driver}; their collaborative interaction is video-recorded via Zoom with a front-facing camera and audio-recorded with each learner wearing a lavalier microphone.}
    \label{fig:collaboration}
\end{figure*}

At the beginning of each pair programming activity, the researchers explained both roles and the expectations for each one. Next, students worked on activities for 35-40 minutes with a randomly assigned partner. During these activities, the teacher and researchers were available to help students with their questions; researchers reminded students to switch roles (and seats) with their partners regularly. 

\subsection{Data Collection}
\begin{figure*}[hbt!]
    \centering
    \includegraphics[width=1\linewidth]{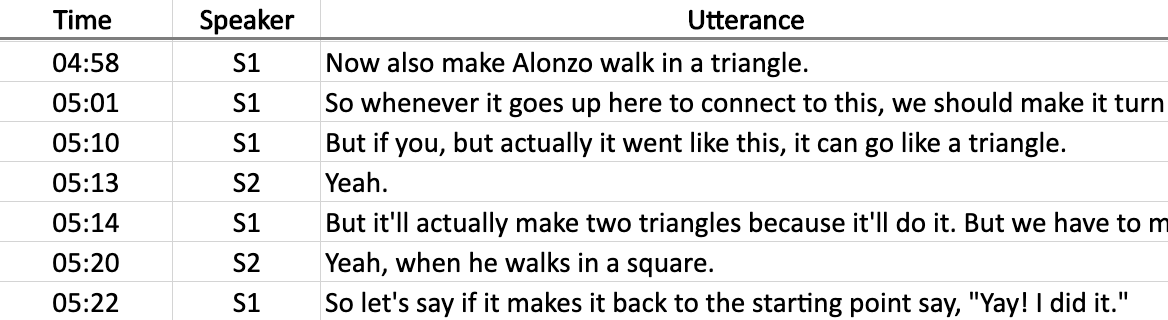}
    \caption{A manually-generated sample transcript excerpt} \label{fig:rev-results}
\end{figure*}

We collected data from learners within a study approved by the institutional review board (IRB) of the University of Florida, and we obtained parental consent and participant assent before the study. The collaborative coding session of each pair was video-recorded via Zoom by the front-facing camera of their laptop (See Figure~\ref{fig:collaboration}); meanwhile, each child wore a headset without active noise canceling, and audio was recorded by the microphone of the headset's built-in digital sound recorders with a sample rate of 48KHz. After the data collection process was finished, the audio recordings were manually transcribed and prepared for tagging (See Figure~\ref{fig:rev-results}).

As shown in Figure~\ref{fig:rev-results}, the transcripts included three pieces of information for each spoken utterance: (1) \textit{Time}, in the form of \textit{min}:\textit{sec}, which indicates the exact starting timestamp for each spoken utterance in the audio recording; (2) \textit{Speaker}, in the form of \textit{S1} (the learner sitting on the left of the video) or \textit{S2} (the learner sitting on the right); and (3) \textit{Utterance}. Each collaborative coding session took around 45 minutes. The corpus included 22 hours and 18 minutes of audio and video recordings, 

\subsection{Manual Annotation of Confusion and Conflict}
In line with prior work on analyzing confusion and conflict dialogues during collaborative learning \citep{goodman2005using, rodriguez2017expressing, tsan2021collaborative}, the process of annotating confusion and conflict was based on textual transcripts, with video used in cases of unresolvable ambiguity within the transcripts. Our dialogue act taxonomy draws upon the exploratory talk framework \citep{mercer2002words} and upon a closely related dialogue act taxonomy by \citet{zakaria2022two} that was designed for elementary school learners’ classroom dialogues. We modified \citet{zakaria2022two}'s dialogue act schema to isolate the exploratory talk moves. We collapsed some of the dialogue act tags into their exploratory talk equivalents (e.g., combining ``Self-Explanation'' with ``Justification'', and ``Suggestion'' with ``Alternative Idea''). We also added a tag to capture utterances that were directed at the agents to separate these from utterances meant for the human partner. Finally, we grouped ``Antagonistic Action'', ``Disagreement / Negative'', and ``Disagreement with Justification'' into one ``Conflict'' category.

To establish the reliability of the dialogue act labeling, two annotators first engaged in a training phase where they collaboratively applied the dialogue act taxonomy and discussed any disagreements. Once training was complete, they independently tagged an overlapping 20\% of the data, reaching a Cohen's kappa score of 0.816, indicating a strong agreement \citep{landis1977measurement}. They then proceeded to divide and tag the remaining data independently. The final dialogue act schema is shown in Table~\ref{tab:annotation-schema} (next page). Among a total of 9,943 transcribed utterances in the corpus, 467 (4.7\%) were labeled as \textbf{Confusion}, 924 (9.3\%) were label as \textbf{Conflict}, and 8,852 (86.0\%) were labeled as \textbf{Other}. 

\begin{landscape}
\begin{table}
\caption{Dialogue Act Schema for Manual Annotation of Confusion and Conflict}
\label{tab:annotation-schema}
\begin{adjustbox}{width = 1.5\textwidth}
\begin{tabular}{@{}cllcc@{}}
\toprule
\textbf{Tag}                    & \multicolumn{1}{c}{\textbf{Description}}                              & \multicolumn{1}{c}{\textbf{Example}}                                                                            & \textbf{Grouped Catogory} & \textbf{Count (Percentage)}   \\ \midrule
Confusion/Seeking Help          & Student directly or indirectly seeking help from partner              & \begin{tabular}[c]{@{}l@{}}``I’m confused'', \\ ``I don't know''\end{tabular}                                       & \textbf{Confusion}                 & 467 (4.7\%)                    \\ \midrule
Antagonistic Action             & Actions or interactions that cause tension                            & \begin{tabular}[c]{@{}l@{}}``You are being ridiculous''\\ ``Stupid''\\ ``You don’t know anything''\end{tabular}       & \multirow{3}{*}{\textbf{Conflict}} & \multirow{3}{*}{924 (9.3\%)}   \\
Disagreement/Negative Feedback  & Disagreement on any opinion/edit                                      & \begin{tabular}[c]{@{}l@{}}``No no no''\\ ``That is wrong''\end{tabular}                                            &                           &                               \\
Disagreement with Justification & Student disagrees, but then explains why                              & ``No because that won’t make it move in a square''                                                                &                           &                               \\ \midrule
Question-Higher Order           & Asks a why question or a question that challenges a partner's idea    & ``Why is he moving like that?''                                                                                   & \multirow{9}{*}{\textbf{Other}}    & \multirow{9}{*}{8,552 (86.0\%)} \\
Question-Other                  & Asks anything other than a why question                               & ``What does that block do?''                                                                                      &                           &                               \\
Acknowledgement                 & Agreement on any opinion/edit                                         & \begin{tabular}[c]{@{}l@{}}``Looks good''\\ ``Good job''\end{tabular}                                               &                           &                               \\
Justification                   & Explain the steps they are taking, or thoughts                        & \begin{tabular}[c]{@{}l@{}}``I did that because of how slow it is''\\``Now we can put this here to make it happen twice''\end{tabular}                                                                         &                           &                               \\
Directive                       & Telling partner to do something                                       & \begin{tabular}[c]{@{}l@{}}``Click that one'' \\ ``Give me the keyboard''\end{tabular}                                                                                                &                           &                               \\
Suggestion/Alternative Idea     & Any suggestions when directly taking to partner (leaving wiggle room) & \begin{tabular}[c]{@{}l@{}}``How about doubling that?'' \\ ``Maybe we should put two of those''\end{tabular}                                                                                      &                           &                               \\
Social                          & Social dialogue                                                       & ``Thanks, we know we are great''                                                                                  &                           &                               \\
Directed at Agents              & It was said to the agent, not the partner                             & \begin{tabular}[c]{@{}l@{}}``Yeah, we know that''\\ ``Shut up''\\ ``Yep, we know we are great partners''\end{tabular} &                           &                               \\
Other                           & Something not covered by any of the other tags                        & {[}Discussion about headphones, sound, volume{]}                                                                &                           &                               \\ \bottomrule
\end{tabular}
\end{adjustbox}
\end{table}
\end{landscape}

\subsection{Automatic Speech Recognition (ASR)}
The manual transcripts were only used for the data annotation process. This study used the automatically recognized transcriptions generated by ASR engines for further multimodal model training and experimental results interpretation. The reason is that practically accurate transcriptions are not available, and intelligent systems that are deployed in real classroom settings has to rely on cloud-based ASR services to transcribe audio recordings in real time \citep{ma2021challenge}.

In this study, each audio segment was automatically transcribed using three popular ASR services: Google Speech-to-text\footnote{\url{https://cloud.google.com/speech-to-text}}, Rev.ai\footnote{\url{https://github.com/revdotcom/revai-python-sdk}}, and OpenAI Whisper\footnote{\url{https://github.com/openai/whisper}}. Google Speech-to-text and Rev.ai are the two widely used commercial ASR engines, and OpenAI Whisper is a popular open-source ASR engine. For each audio segment, each of the ASR services generated a transcribed utterance along with a confidence score for each transcribed word. To assess accuracy of the automatic transcription, we used the manual transcription as ``gold-standard''. We used the metric of word error rate ($WER$), given by $WER = (S + D + I ) / N$, where $S$ is the number of substitutions, $D$ is the number of deletions, $I$ is the number of insertions, and $N$ is the number of words in the human transcript. Table~\ref{tab:wer} shows the comparison results of the three ASR engines.

\begin{table}[htbp]
\centering
\caption{WER for Each ASR Engine}
\label{tab:wer}
\begin{tabular}{@{}lcccc@{}}
\toprule
\multicolumn{1}{c}{}  & Mean & SD   & Min  & Max  \\ \midrule
Google Speech-to-Text & 0.78 & 0.24 & 0.63 & 1.58 \\
Rev.ai                & 0.93 & 0.46 & 0.69 & 2.41 \\
OpenAI Whisper        & 0.84 & 0.54 & 0.55 & 5.38 \\ \bottomrule
\end{tabular}
\end{table}

Compared among the three popular ASR engines, we found that Google Speech-to-Text achieved the most robust performance in the face of substantial audio noise, with the lowest mean WER score of 0.78 and the lowest standard deviation of 0.24. Although OpenAI Whisper achieved the lowest minimum WER score of 0.55, its performance is not robust when the level of audio noise increases, indicated by the highest maximum WER score of 5.38. Finally, we selected the transcriptions generated from Google Speech-To-Text for further multimodal model training and experimental results interpretation.

\section{Multimodal Features}\label{sec4}
In this study, we extracted multimodal features from each spoken utterance and its corresponding audio and video during elementary school learners' collaborative dialogues. The organization of this section is as follows: we first describe the preprocessing of the multimodal data for feature extraction in Subsection \ref{4.1}; next, we introduce the feature extraction process from language (Subsection \ref{4.2}), audio (Subsection \ref{4.3}), and video (Subsection \ref{4.4}) modalities. Last, we describe the feature postprocessing and embedding subnetworks in Subsection \ref{4.5}, \ref{4.6} that embed the extracted audio- and video-derived features for model training.

\subsection{Data Preprocessing} \label{4.1}
With the help of the starting timestamp of each utterance, we used \textit{pydub.AudioSegment}, a function in the pydub\footnote{\url{https://github.com/jiaaro/pydub}} library, to read audio files and extract audio segments with given starting timestamps; similarly, we used \textit{cv2.VideoWriter}, a function in the OpenCV\footnote{\url{https://github.com/opencv/opencv-python}} library, to automatically extract video segments of each spoken utterance. Before extracting features for each spoken utterance from transcriptions generated by Google ASR, we used NLTK\footnote{\url{https://www.nltk.org/}}, a natural lexical processing tookit, to apply a few text preprocessing steps. These steps included (1) removing extra white spaces and spacial characters (e.g., ``[]'',``\{\}''), (2) expanding contractions, (3) tokenization, and (4) lower casing.

\subsection{Language-derived Features} \label{4.2}
Language-derived features have been widely used to model collaborative problem-solving skills and predict collaborative task performance \citep{reilly2019predicting,vrzakova2020focused}. In this study, we extracted the following three language-derived features from each utterance transcribed by the Google ASR engine:
\begin{enumerate}
    \item Term frequency-inverse document frequency (TF-IDF). TF-IDF is a widely used statistical method that provides insights into words and phrases that are statistically more important \citep{kusner2015word}. In this study, we used TF-IDF to measure how important terms and phrases are within an utterance relative to the whole dataset.
    \item RoBERTa. In this study, we fine-tuned the pre-trained RoBERTa model to generate sentence embeddings. BERT is a language model trained on a large amount of data (e.g., texts from Wikipedia and books) in a self-supervised way \citep{devlin2018bert}, and RoBERTa is a variant of BERT trained on longer sequences. RoBERTa dynamically changes the masking pattern applied to the training data and has outperformed BERT on a series of language processing tasks, such as machine translation and question answering \cite{liu2019roberta}. We used the HuggingFace xlm-roberta-based\footnote{\url{https://huggingface.co/xlm-roberta-base}} model to generate a fixed-dimensional (in this study, we set the encoding dimension $d=768$) embedding for each utterance.
    \item Sentiment. Sentiment indicates the attitudes of speakers (e.g., positive, neutral, and negative). In this study, we used the HuggingFace twitter-roberta-base\footnote{\url{https://huggingface.co/cardiffnlp/twitter-roberta-base-sentiment}} model to generate the sentiment of each transcribed utterance. This model is a roBERTa-base model trained on around 58 million tweets and then fine-tuned for sentiment analysis with the TweetEval benchmark \citep{barbieri2020tweeteval}.
\end{enumerate}

\subsection{Audio-derived Features} \label{4.3}
Simple acoustic-prosodic features (e.g., sound level, synchrony in the rise and fall of the pitch) derived from audio have been used for analyzing collaborative learning processes, such as estimating group performance on solving open-ended tasks \citep{subburaj2020multimodal} and predicting rapport building \citep{lubold2014acoustic}. In our study, we extracted audio-derived features on the corresponding audio segment for each utterance. 

We used openSMILE v2.2, an open-source toolkit for automatic acoustic feature extraction\footnote{\url{https://audeering.github.io/opensmile-python}}, with the eGeMAPSv02 feature configuration \citep{eyben2015geneva}, which is a popular acoustic feature set used for automatic voice analysis \citep{song2021frustration} and speech emotion recognition \citep{mohamad2019shemo}. This configuration extracts low-level descriptors (LLD) from 20-ms windows of the speech signal with 10-ms window shifts, which is a common configuration that has been used in many speech and audio processing applications \citep{eyben2013recent}. From all LLDs extracted by openSMILE, we select the following five categories of acoustic-prosodic features:
\begin{enumerate}
    \item Loudness (Intensity). Loudness measures the energy level of the audio signal in the time domain. For each audio frame, 11 loudness-related features were extracted, including the mean and the standard deviation loudness values, as well as the number of loudness peaks per second.
    \item Pitch. Pitch measures the frequency scale of a signal in the frequency domain. For each audio frame, ten pitch-related features were extracted, including the fundamental frequency, as well as the mean and the standard deviation pitch values.
    \item Shimmer. Shimmer measures how quickly the loudness of the signal is changing, computed as the average of the relative peak amplitude differences over frames. For each audio frame, two shimmer-related features were extracted: the mean and the standard deviation shimmer values.
    \item Jitter. Jitter measures how quickly the frequency of the signal is changing, computed as the average of the absolute pitch differences over frames. For each audio frame, two jitter-related features were extracted: the mean and the standard deviation jitter values.
    \item Mel-Frequency Cepstral Coefficients (MFCCs). MFCCs measure the shape of the signal’s short-term spectrum. For each audio frame, 16 MFCC-related features were extracted, including the mean and the standard deviation values of lower-order MFCC 1-4.
\end{enumerate}

\subsection{Visual Features} \label{4.4}
A variety of features generated from the video modality have been investigated in prior literature to model collaborative problem solving. For example, eye gaze has proven effective in evaluating learners' attentiveness \citep{schneider2018leveraging,huang2019identifying} and learning performance \citep{celepkolu2018predicting,rajendran2018predicting}; head pose has also been used for assessing learners' collaborative problem solving competence \citep{cukurova2020modelling}; facial action units (AUs) shown to be predictive of individual learners' frustration and engagement level \citep{grafsgaard2013automatically} and interaction level during collaborative learning \citep{malmberg2019going}. Body pose has been used for analyzing learners' engagement level \citep{radu2020relationships}, modeling collaborative problem solving competence \citep{cukurova2020modelling}, and predicting learners' satisfaction levels toward their partners \citep{ma2023automatically}. In our study, video-derived features were extracted from the corresponding raw video segment of each utterance.

We used OpenFace\footnote{\url{https://github.com/TadasBaltrusaitis/OpenFace}} v2.0 facial behavior analysis toolkit to automatically extract video-based features. We used the \textit{multiple faces} mode to extract the following three categories of visual features:
\begin{enumerate}
    \item Eye Gaze Direction. Eye gaze direction refers to the direction in which an eye looks. For each detected face per video frame, 8 eye gaze direction-related features were extracted.
    \item Head Pose. Head pose refers to head position and direction. For each detected face per video frame, 6 head-related features were extracted with OpenFace, including three head position vectors (x direction, y direction, and z direction) representing the location of the head with respect to the camera in millimeters, and three head direction vectors in radius with respect to the camera. 
    \item Facial Action Units (AUs). Facial AUs refer to the movements of an individual’s facial muscles and are commonly used to describe human facial expressions. Movements of facial muscles are taxonomized according to their appearance on the face by Facial Action Coding System\footnote{\url{https://www.cs.cmu.edu/~face/facs.htm}}. Research has shown that specific types of AUs could be related to learners' different cognitive states (e.g., Brow Lower (AU4) and eyelid tightening (AU7) are associated with confusion \citep{grafsgaard2013automatically}). In our study, for each detected face per video frame, 35 facial AU-related features were automatically extracted with OpenFace, including 17 facial AU intensity features (how intense is the AU, ranging from 0 to 5), and 18 facial AU presence features (if the AU is visible in the face, 0-absence or 1-presence).
\end{enumerate}

\subsection{Feature Postprocessing} \label{4.5}
After we used OpenFace to automatically generate video-based features, we manually corrected the output features if more than two faces were detected in the video frame or two learners switched seats during a session. For example, other non-related faces could be captured by the camera when students from other pairs were seated in the background; the detected visual features and learner identities would be mismatched after two learners switched seats during a session. In these cases, we manually removed the features related to these detected non-related faces after the automatic video-based feature extraction process.

Because features from each modality do not share the same range of values, it negatively impacts the machine learning models' training stability since gradients will oscillate back and forth and take a long time before they can finally find their way to the global/local minimum \citep{kahou2016emonets}. To ensure the training stability and model performance, we z-scored and normalized all features to the range of -1 to 1 by subtracting their mean value and dividing by their standard deviation. We used the combined statistics of the entire training set to capture the overall distribution and variability of the multimodal features. Table~\ref{tab:features} lists the details of the multimodal features extracted and investigated in this study and their corresponding vector dimensions. Because audio- and video-derived features were initially extracted at millisecond-level, in the next subsection, we used transformer-based networks to embed audio- and video-derived features into lower dimensional space with fixed-length, so as to prepare multimodal features for further model training.

\begin{table}
    \caption[Multimodal features and dimensions.]{Feature categories and dimensions extracted from each modality. The dimensions for language-derived features are utterance-level; the dimensions for audio- and video-derived features are frame-level, and are later embedded by transformer-based embedding networks (See Subsection~\ref{4.6}) to form utterance-level feature representations.}
    \label{tab:features}
    \centering
    \begin{threeparttable}[b]
    \begin{tabular}{ccc}
        \toprule
        Modality         & Feature Name & Feature Dimension \\ \toprule
        \multirow{3}{*}{Language} & TF*IDF           & 1,909                          \\
                                  & RoBERTa          & 768                         \\
                                  & Sentiment            & 3                     \\ \midrule
        \multirow{6}{*}{Audio}    & Loudness             & 11              \\
                                  & Pitch                & 10              \\
                                  & Shimmer              & 2             \\
                                  & Jitter               & 2              \\
                                  & MFCCs                & 16            \\
                                  & Wav2Vec              & 768             \\ \midrule
        \multirow{3}{*}{Video}    & Eye Gaze             & 112                        \\
                                  & Head Pose            & 6                       \\
                                  & Facial AUs           & 35                      \\ \bottomrule
    \end{tabular}
    \end{threeparttable}
\end{table}

\subsection{Feature Embedding Subnetworks for Audio- and Video-derived Features} \label{4.6}
In this study, audio- and video-derived features were initially extracted at frame-level with automatic feature extraction toolkits (openSMILE for audio and OpenFace for video). To form utterance-level audio and video features, it is necessary to embed the sequence of frame-level audio- and video-derived features into a fixed-length feature representation. Inspired by the superior performance of transformer models in tasks of time series modeling \citep{zerveas2021transformer,faouzi2020pyts} and forecasting \citep{zhou2021informer,zeng2023transformers}, we used \textit{longFormer} \citep{beltagy2020longformer}, a type of transformer-based neural network architecture designed for handling long-range dependencies in sequences. Transformers, introduced by \citet{vaswani2017attention}, have been highly successful in various natural language processing tasks. However, a limitation of the original transformer architecture lies in its incapablity of handling long sequences (\textit{max\_sequence} = 512) because of its quadratic complexity with respect to the sequence length. LongFormer architecture introduces sliding window attention that allows the model to attend to only a subset of tokens within a window, reducing the overall computational complexity. This design makes LongFormer well-suited for processing longer sequences (\textit{max\_sequence} = 4096).

\begin{figure*}[!tbp]
    \centering
    \includegraphics[width=.85\textwidth]{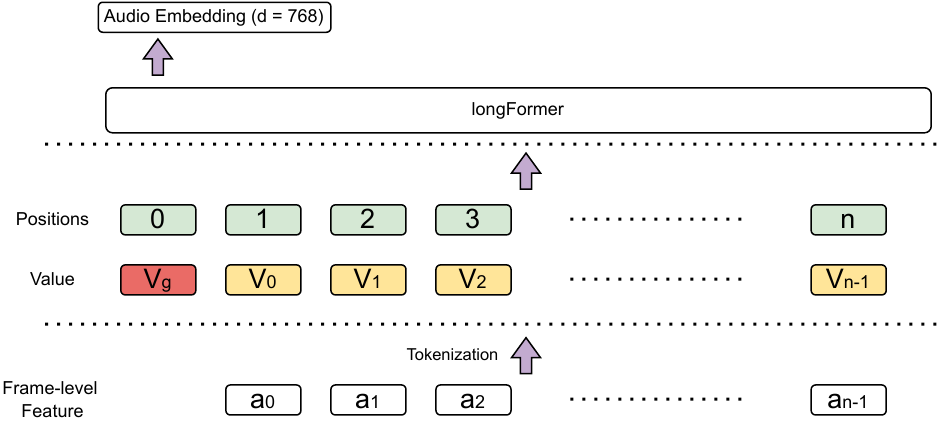}
    \caption{Transformer-based feature embedding subnetwork.}
    \label{fig:longformer}
\end{figure*}

Figure~\ref{fig:longformer} shows a high-level overview of the longFormer-based feature embedding network for encoding frame-level audio- and video- derived features into a fixed-length feature embedding. Taking a frame-level audio-derived feature sequence $X_{a}$ as an example. Typical transformer tokenization processes for texts include word tokenization (e.g., break the text into words and subword units), adding special tokens (e.g., ``CLS'' token marking the beginning of the sequence), and positional encoding, which adds the order of tokens to the token embeddings. Our embedding scheme mainly differs with typical tokenization scheme in two ways. Giving a frame-level audio-derived feature sequence $X_{a} = [a_0, a_1, ..., a_{n-1}]$, first, work tokenization is replaced by a linear layer encoding the feature value (\textit{dim}=1) into a value embedding $V_{a}$ (\textit{dim}=64); second, special tokens like ``CLS'' replaced by a global attention token $V_g$ prepended to the value embedding sequence that is meant to capture a global representation of the entire sequence. After tokenization, the value embeddings and positional embeddings are summed to form input embedding for longFormer. Final audio embedding(\textit{dim}=768) is generated from the output of final layer of longFormer.

\section{Unimodal Modeling of Confusion and Conflict} \label{sec5}
Identifying predictive unimodal features helps with filtering out noisy features that potentially have a low correlation with confusion and conflict. Therefore, we first built several supervised unimodal models trained on linguistic, acoustic, and visual features to detect confusion and conflict.

\subsection{Experimental Setup}
We trained both classical machine learning classifiers (e.g., Support Vector Machine, Random Forest) and neural-network-based Multi-layer Perceptron (MLP) classifiers with each of the unimodal features to identify predictive unimodal features for the task of confusion and conflict detection. We chose Multilayer Perceptron (MLP) as the classifier because it is flexible to adjust its number of hidden layers and activation functions to learn the complex patterns in the unimodal data. The MLP classifier contains two feed-forward layers with an embedding size of 128 and an activation function of \textit{ReLU}, followed by two dropout layers with a rate of 0.5 were added to each linear layer, respectively to alleviate over-fitting. The \textit{Softmax} activation function was used in the last output layer to generate multi-classification results, namely \textbf{Confusion}, \textbf{Conflict}, and \textbf{Other}. Since the label distribution of \textbf{Confusion} (4.7\%) and \textbf{Conflict} (9.3\%) within our corpus are highly imbalanced, the classifiers would have poor performance detecting confusion and conflict samples, as they would have been trained mostly on data with \textbf{Other} samples. To mitigate the effect of the label imbalance, we used Synthetic Minority Oversampling Technique (SMOTE) \citep{chawla2002smote} to synthesize new confusion and conflict samples in training datasets. SMOTE is a widely used data augmentation technique that works by selecting examples close to the feature space, drawing a line between the examples in the feature space, and drawing a new sample at a point along that line \citep{mathew2015kernel}. This approach ensures an approximately even number of training samples for three categories. SMOTE was only performed on the training set, and class distributions for the validation and testing sets were left unchanged. In SMOTE, k\_neighbors represents the number of nearest neighbors to be considered when generating synthetic samples. The choice of k can impact the quality and effectiveness of the synthetic samples. In this study, we used the default k\_neighbors of 5 for upsampling minority classes. We experimented with different k values and the results were similar.

We implemented the Python code for our prediction models in Keras with a Tensorflow backend. We conducted five-fold subject-independent cross-validation \citep{gholamiangonabadi2020deep}, in which the datasets were split into the training and testing sets by subject. We chose subject-independent cross-validation over traditional cross-validation because samples that belong to the same subject are likely to be related to each other, due to underlying environmental, biological and demographics factors \citep{dehghani2019subject}, which may lead to artificially increasing the model performance. We chose the fold number \textit{k}=5 because the sample size of our dataset is relatively small (N=38); thus, using a higher number of folds (e.g., ten folds) could lead to each fold having a very limited number of samples, which might result in unstable performance estimates. Therefore, we selected five-fold to ensure that each fold has a reasonable number of samples for evaluation. The fine-tuning process of BERT (Bidirectional Encoder Representations from Transformers) involves taking a pre-trained BERT model and further training it on a specific downstream task using task-specific labeled data. Fine-tuning also involves optimizing hyperparameters, such as learning rate, batch size, and number of training epochs. These hyperparameters can significantly impact the performance of the fine-tuned model, and tuning them is crucial to achieve the best results on the specific task. We used Adam optimizer \citep{kingma2014adam} with the learning rate of $1 \times e^{-3}$ to train the prediction model, which was trained for up to 100 epochs.

Evaluation Metrics
\begin{enumerate}
    \item Accuracy. Accuracy measures the overall proportion of correctly classified instances of Confusion, Conflict, and Other. However, since Confusion and Conflict samples only account for 4.7\% and 9.3\% of our dataset, respectively, solely relying on accuracy might be misleading because a model that predicts Other (the majority class) for all instances can still achieve high accuracy. 
    \item Precision. Precision is the ratio of true positives to the total number of samples that are classified as positives. In this study, the precision represents the ability of a model to correctly classify confusion or conflict samples. High precision indicates that when the model predicts a confusion or conflict sample, it is more likely to be correct.
    \item Recall. Recall is the ratio of samples that are classified as positives to the total number of actual positives. In this study, recall represents the ability of a model to classify all true confusion or conflict samples. Higher recall indicates that the model can detect a greater portion of true confusion or conflict samples.
    \item Macro F-1 score. Combining precision and recall, the F-1 score is the harmonic mean of precision and recall. It balances both precision and recall and provides a single metric that considers both false positives and false negatives. In this study, we evaluated the overall performance of the trained classifier with the macro F-1 score, which is calculated by averaging the sum of the F-1 score of each of the three classes: Confusion, Conflict, and Other. Another commonly used F-1 score metric is the weighted F-1 score, which takes the class frequencies into account. Since confusion and conflict samples belong to the minority classes that appear in the dataset, while they are more important than other samples that belong to the majority class, the macro F-1 score is more appropriate than the weighted F-1 score because it considers each class equally important. The weighted F-1 score, on the contrary, favors the majority class. In adition, we also use precision and recall provides additional information. The context of collaborative dialogue may shift the cost of false negatives versus false positives, so these additional scores allow us to weigh each case.
\end{enumerate}

\begin{table*}[hbtp]
\caption[Results for best-performing unimodal features.]{Results for selecting best-performing unimodal features. SMOTE was applied on the training set while the label distribution for the testing set remained imbalanced:  Confusion (4.7\%), Conflict (9.3\%), and Other (86.0\%). P: Precision, R: Recall, F: F-1 Score, A: Overall Accuracy. MF: Macro F-1 Score.}
\label{tab:feature-selection}
\begin{adjustbox}{width = 1\textwidth}
\begin{tabular}{@{}ccccccccccccc@{}}
\toprule
\textbf{} & \textbf{} & \multicolumn{3}{c}{Confusion} & \multicolumn{3}{c}{Conflict} & \multicolumn{3}{c}{Other} & \textbf{} & \textbf{} \\
Modality & Unimodal Features & P & R & F & P & R & F & P & R & F & A  & MF\\ \midrule
\multirow{3}{*}{Language} & TF*IDF & 0.33 & 0.41 & 0.36 & 0.19 & 0.34 & 0.28 & 0.84 & 0.70 & 0.76 & 0.65 & 0.47\\
 & RoBERTA & 0.25 & 0.68 & 0.37 & 0.37 & 0.66 & 0.43 & 0.89 & 0.81 & 0.85 & 0.74 & 0.55 \\
 & Sentiment & 0.19 & 0.64 & 0.30 & 0.20 & 0.59 & 0.31 & 0.85 & 0.50 & 0.64 & 0.61 & 0.42 \\
\multirow{6}{*}{Audio} & Loudness & 0.13 & 0.10 & 0.11 & 0.17 & 0.26 & 0.20 & 0.86 & 0.83 & 0.85 & 0.62 & 0.39 \\
 & Pitch & 0.18 & 0.10 & 0.13 & 0.10 & 0.17 & 0.14 & 0.81 & 0.93 & 0.86 & 0.63 & 0.38 \\
 & Shimmer & 0.15 & 0.23 & 0.19 & 0.21 & 0.26 & 0.24 & 0.86 & 0.64 & 0.73 & 0.60 & 0.39 \\
 & Jitter & 0.17 & 0.30 & 0.24 & 0.23 & 0.30 & 0.26 & 0.83 & 0.76 & 0.79 & 0.61 & 0.40 \\
 & MFCCs & 0.18 & 0.34 & 0.25 & 0.26 & 0.37 & 0.27 & 0.90 & 0.70 & 0.80 & 0.66 & 0.44 \\
 & Wav2Vec & 0.25 & 0.45 & 0.32 & 0.27 & 0.49 & 0.36 & 0.87 & 0.88 & 0.88 & 0.70 & 0.52 \\
\multirow{3}{*}{Video} & Eye Gaze & 0.17 & 0.32 & 0.23 & 0.33 & 0.24 & 0.29 & 0.85 & 0.82 & 0.83 & 0.68 & 0.45 \\
 & Head Pose & 0.19 & 0.27 & 0.22 & 0.31 & 0.21 & 0.27 & 0.88 & 0.85 & 0.86 & 0.65 & 0.42 \\
 & Facial AUs & 0.40 & 0.54 & 0.46 & 0.31 & 0.39 & 0.32 & 0.89 & 0.86 & 0.88 & 0.73 & 0.55 \\ \bottomrule
\end{tabular}
\end{adjustbox}
\end{table*}

The experimental results of unimodal modeling of confusion and conflict from Table~\ref{tab:feature-selection} showed that the best-performing unimodal features in each modality: fine-tuned RoBERTa, Wav2Vec, and facial AUs in the language, audio, and video modality, respectively. In language derived features, semantic features generally outperformed TF*IDF. The F-1 scores for both \textbf{Confusion} and \textbf{Conflict} classes were improved as more semantic information was included, and Fine-tuned RoBERta performed the best with the highest F-1 scores of 0.37 for \textbf{Confusion} and 0.43 for \textbf{Conflict}, and the highest overall accuracy of 0.74. Interestingly for \textbf{Conflict}, the statistical feature yielded comparable results (F-1 score of 0.36) RoBERTa (F-1 score of 0.37). The reason may be that the confusion cases in our corpus are relatively simple, mainly including cases like: ``I don’t know.'', ``We need help.'', and ``I’m confused.''; statistical methods which compute the frequency of signal words may be simple yet powerful enough to detect those simple confusion cases. In audio-derived features, only audio embeddings generated by pre-trained Wav2Vec provided the best performance in detecting \textbf{Conflict}, with the highest F-1 score of 0.36. All acoustic-prosodic features investigated in this study yielded relatively poor performance in detecting \textbf{Confusion}, with none of the F-1 scores of the unimodal models trained with these features beyond 0.20. In video-derived features, facial AUs provided the best performance in detecting \textbf{Confusion}, with the highest F-1 score of 0.46. All visual features performed relatively poor in detecting \textbf{Conflict}, and none of their F-1 scores was above 0.30. The unimodal model trained with facial AUs also had the highest overall accuracy of 0.73.

\section{Multimodal Modeling of Confusion and Conflict} \label{sec6}
Next, we examined the performance of combining these predictive unimodal features into a multimodal model. Prior research on modeling collaborative learning has shown that combining information from language, audio, and video modalities can increase model performance as they measure unique aspects of the collaboration \citep{stewart2021multimodal}. In this study, we built supervised multimodal models trained on subsets of the combination of linguistic, acoustic, and visual features to examine the performance of combining these predictive unimodal features identified in Section~\ref{sec5} into a multimodal model. Followed the multimodal feature fusion methods widely adopted in prior works, we compared the performance of a series of multimodal feature fusion baselines (i.e., early fusion and late fusion \citep{dong2009advances}) as well as state-of-the-art neural-network-based methods (tensor fusion network \citep{zadeh2017tensor} and cross-attention fusion network \citep{tsai2019multimodal}).

\begin{figure*}[!tbp]
    \centering
    \includegraphics[width=.55\textwidth]{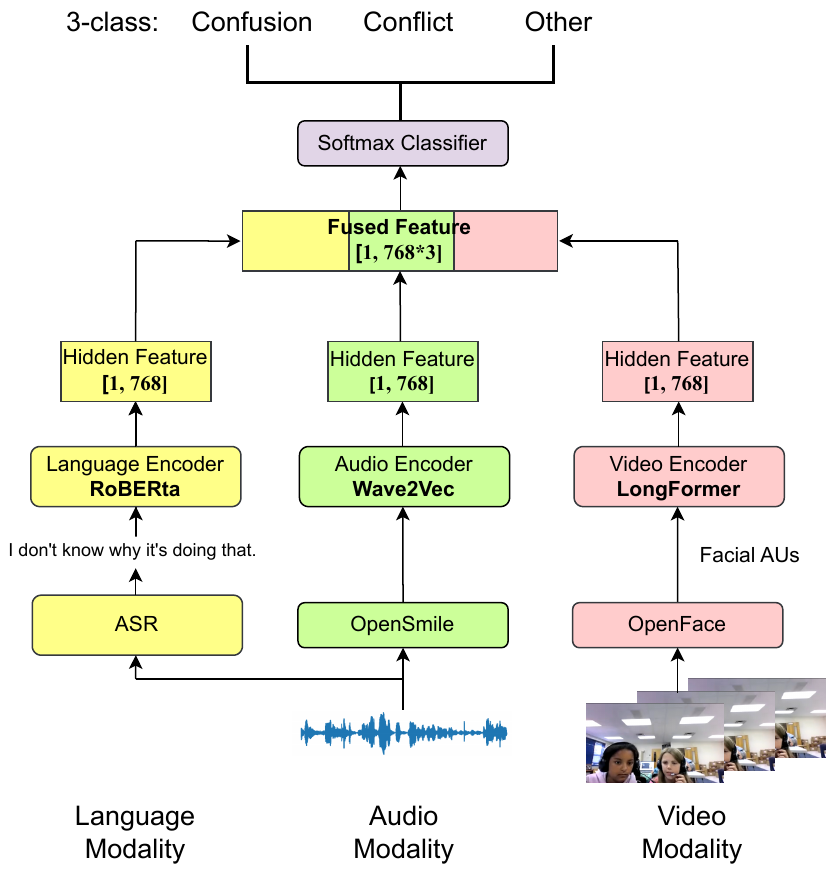}
    \caption{Overview of the multimodal architecture of the confusion and conflict moments detection model with early fusion. The other multimodal models followed the same structure with a subset of the modalities.}
    \label{fig:early-fusion}
\end{figure*}

Figure~\ref{fig:early-fusion} shows the overview of the multimodal architecture of the confusion and conflict detection model with early fusion. We followed the early fusion strategy to combine these unimodal features into a single multimodal feature. The combined multimodal feature is fed into a Softmax classifier to predict the probability of the past moment belonging to each of the three classes, i.e., Confusion, Conflict, or Other. Before concatenating unimodal feature vectors into a single multimodal feature vector, we applied $z$-score normalization to all the features by subtracting their mean value and dividing them by their standard deviation. For late fusion, three parallel models were trained separately based on predictive features from each of the three modalities. Then, the classification scores from the three modalities were averaged to compute the final estimated satisfaction score.

\begin{figure*}[!tbp]
    \centering
    \includegraphics[width=.55\textwidth]{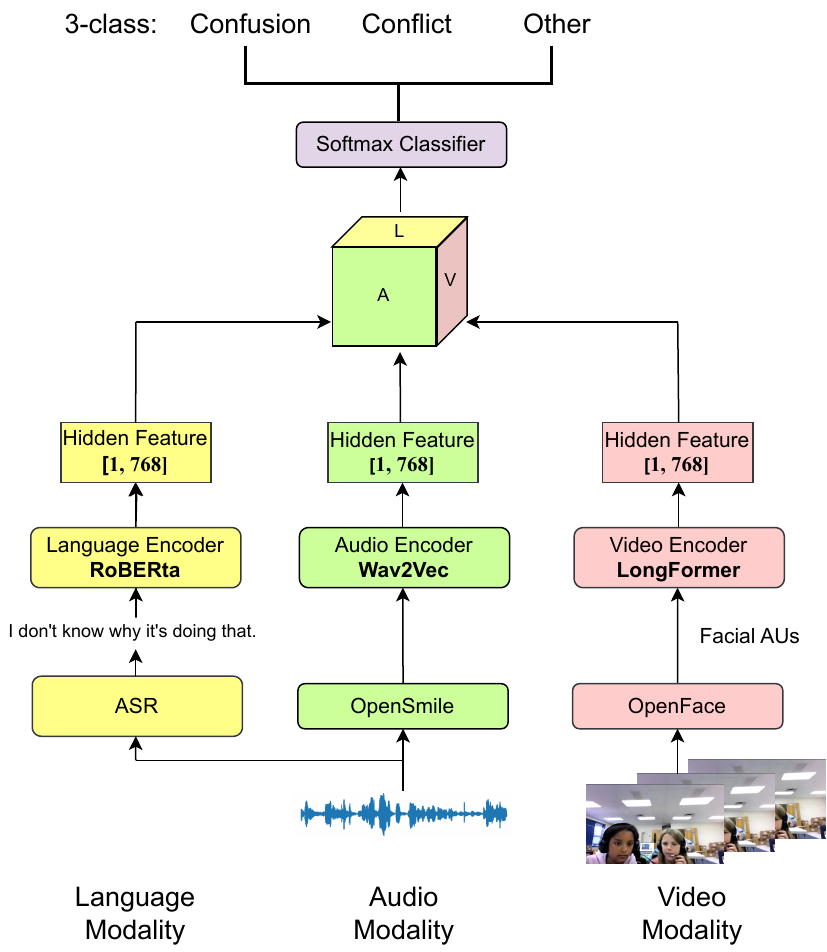}
    \caption{Overview of the tensor fusion-based multimodal architecture.}
    \label{fig:tensor-fusion}
\end{figure*}

To compare across different feature fusion methods, we also experimented with two more recent neural-network-based fusion networks, namely tensor fusion network and cross-attention network. Tensor fusion transforms multimodal features into a 3D feature tensor, while cross-attention fusion uses a shared transformer encoder to attend to different modalities. The tensor fusion network has three main layers: fusion, encoding, and prediction. The fusion layer combines the different input data modalities, which may have different dimensions and feature representations, into a unified tensor representation. The encoding layer then maps this tensor representation to a lower-dimensional latent space, which can be processed by standard neural network layers. Finally, the layer component outputs the result based on the encoded tensor representation. In this study, we only used the tensor fusion layer to generate a 3D cube multimodal feature tensor of all possible interactions among three modalities (see Figure~\ref{fig:tensor-fusion}), while early fusion can be seen as a special case of tensor fusion with only unimodal interactions that generate a 2D multimodal vector.

\begin{figure*}[!tbp]
    \centering
    \includegraphics[width=.55\textwidth]{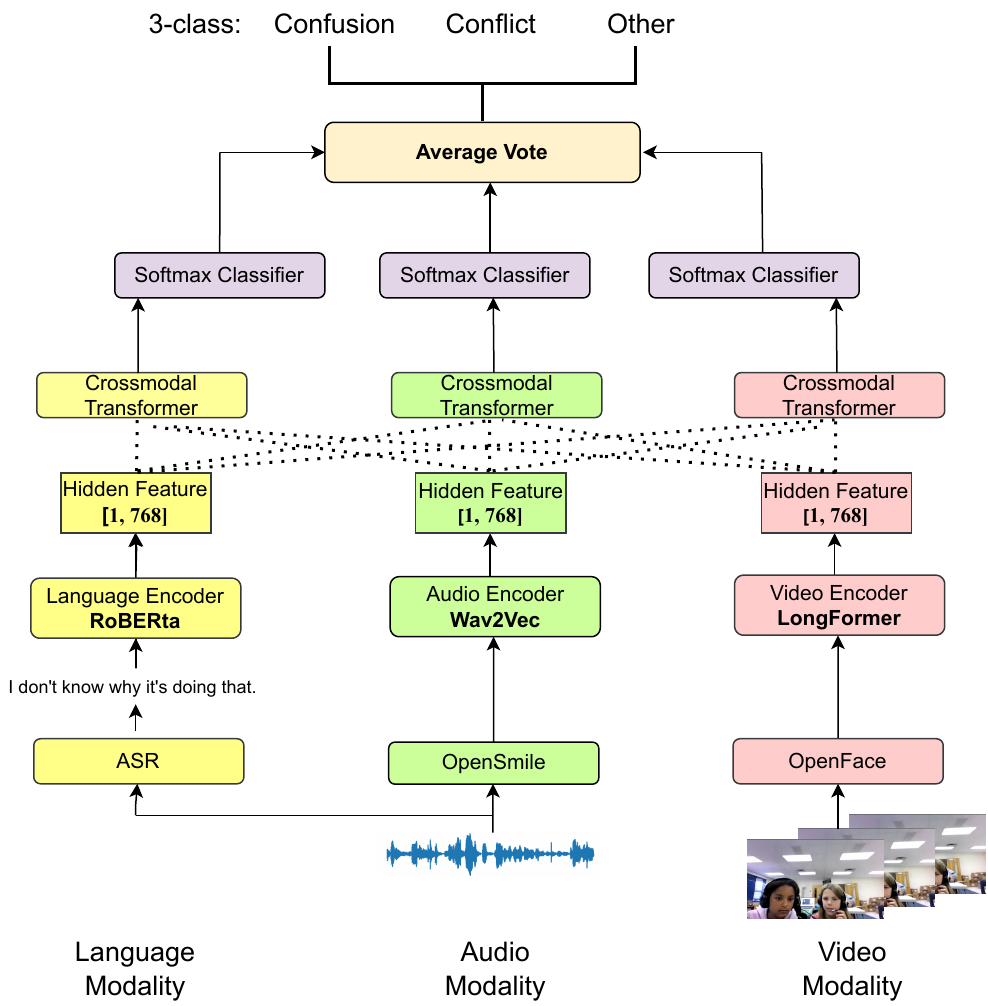}
    \caption{Overview of the cross-attention multimodal architecture.}
    \label{fig:cross-fusion}
\end{figure*}

Another fusion method based on an attention mechanism has gained research interest recently. 
\citet{tsai2019multimodal} proposed the cross-attention mechanism, where instead of early-fusing features from multiple modalities, the cross-attention mechanism enables a feature from one modality to enrich its information by searching for the most relevant feature in the other modality. In order to learn the correlations and dependencies between the two modalities, the cross-attention mechanism provides a latent adaptation across modalities. The feature sequence in one modality is enriched by feature sequence from the other modality by first searching the most relevant information from modality with computing a scoring matrix, whose $(i, j)$-th element measures the attention given by the information from the $i$-th time step from one modality and the $j$-th time step from the other modality. Inspired by the cross-attention mechanism, this subsection uses crossmodal transformer-based fusion (See Figure~\ref{fig:cross-fusion}), in which the core is a deep stacking of several cross-attention + feedforward blocks.

We implemented tensor fusion based on open-source code.\footnote{\url{https://github.com/A2Zadeh/CMU-MultimodalSDK/tree/master/mmsdk/mmmodelsdk/fusion/tensor_fusion}} We used a tensor fusion layer to generate multimodal feature tensor from predictive unimodal features from lexical, audio, and video modality. We implemented cross-attention based on open-source code.\footnote{\url{https://github.com/yaohungt/Multimodal-Transformer}}. The training settings were kept the same as in Section~\ref{sec5}, where SMOTE was applied on the training set to mitigate the class imbalance issue, and class distributions for the validation and testing sets were left unchanged. We conducted five-fold cross-validation to train and validate the models. The network weights were learned with Adam optimizer with a learning rate of 0.0001, dropout rate of 0.2, and l2 regularizer of 0.01 were used to avoid over-fitting. The model was trained up to 50 epochs, and a mini-batch of 16 samples was used for training at each iteration, stopping early if validation loss did not decrease for 15 epochs. We evaluated the performance of the trained models using overall accuracy and F-1 scores combined from precision and recall for each one of the three classes.

Table~\ref{tab:rq1} shows the confusion and conflict moments classification performance for models trained on each of the single unimodal features. The experimental results showed that the Cross-Attention Fusion approach performed the best, with higher overall accuracy and F-1 scores for classifying confusion and conflict moments in the dataset. These results indicated that neural-network-based multimodal fusion methods have a stronger capability for capturing the correlations and dependencies between modalities compared to non-neural-network-based fusion methods.

\begin{table*}[hbtp]
\centering
\caption[Performance of different multimodal models.]{Performance of different multimodal data fusion models. Two baseline models are non-neural-network-based, while Tensor Fusion and Cross-Attention Fusion are neural-network-based. SMOTE was applied on the training set while the label distribution for the testing set remained imbalanced:  Confusion (4.7\%), Conflict (9.3\%), and Other (86.0\%). P: Precision, R: Recall, F: F-1 Score, A: Overall Accuracy. MF: Macro F-1 Score.}
\label{tab:rq1}
\begin{adjustbox}{width = 1\textwidth}
\begin{tabular}{@{}lccccccccccc@{}}
\toprule
\multicolumn{1}{c}{}                & \multicolumn{3}{c}{Confusion} & \multicolumn{3}{c}{Conflict} & \multicolumn{3}{c}{Other} &   \\
\multicolumn{1}{c}{Fusion Approach} & P        & R        & F       & P        & R       & F       & P       & R      & F      & A & MF\\ \midrule
Early Fusion (Baseline 1)                      & 0.41     & 0.54    & 0.48     & 0.39    &  0.61   & 0.53    & 0.89   & 0.84   & 0.86   & 0.74 & 0.62\\
Late Fusion (Baseline 2)                       & 0.35     & 0.49    & 0.43     & 0.34    & 0.47    & 0.41    & 0.87   & 0.80   & 0.83   & 0.72 & 0.56\\
Tensor Fusion                       & 0.45     & 0.58     & 0.51    & 0.43     & 0.64    & 0.54    & 0.87    & 0.88   & 0.88   & 0.80  & 0.64\\
Cross-Attention + Early Fusion      & 0.53     & 0.61     & 0.57    & 0.49     & 0.65    & 0.56    & 0.88    & 0.85   & 0.86   & 0.83   & 0.66\\
Cross-Attention + Late Fusion & 0.51  & 0.69     & 0.59      & 0.54    & 0.72    & 0.62    & 0.93   & 0.88  & 0.91  & 0.87 & 0.70 \\ \bottomrule
\end{tabular}
\end{adjustbox}
\end{table*}

The experimental results showed that the cross-attention fusion approach performed the best, with higher overall accuracy and F-1 scores for classifying confusion and conflict moments in the dataset. These results indicated that neural-network-based multimodal fusion methods have a stronger capability for capturing the correlations and dependencies between modalities compared to non-neural-network-based fusion methods.

\section{Discussion}\label{sec8}
Collaborative learning provides numerous benefits to learners; yet, the naturally emerging moments of confusion and conflict during the collaboration process have the potential to generate frustration and significantly impede learners' group performance. Our ultimate goal is to build real-time intelligent systems that detect the confusion and conflict moments during collaborative learning and provide adaptive support for learners to work through them. In this paper, we reported on studies investigating the automatic detection of confusion and conflict using linguistic, acoustic, and visual features from learners' interactions during collaborative learning activities. This section discusses the results with respect to our two research questions, as well as the implications of comparisons among the performance of different model architectures.

\subsection{Main Findings}
Our first research question asked: \textit{What are the predictive unimodal features to detect confusion and conflict during learners’ collaborative interactions?} This study identified the most predictive unimodal feature to detect confusion and conflict during learners' collaborative interactions, which informs the design of the protocol to ensure that high-quality multimodal data is collected. The results showed that the simple statistical method (TF*IDF) yielded comparable performance with sentence embeddings generated from RoBERTa-based transformer models for detecting Confusion, but worse performance for detecting Conflict. This is likely because the confusion samples in our corpus frequently include simple utterances like: ``I don’t know,'' ``We need help,'' and ``I’m confused.'' However, TF*IDF performed worse than RoBERTa for detecting conflict. The potential reason may be that the ways in which conflicts emerge during discourse are more diverse. For example, \citet{tsan2021collaborative} summarized multiple types of conflict dialogues during upper elementary school students' pair programming activities, including ``task conflict'' that relates to their coding tasks, ``control conflict'' that involves the control of the computer, and ``partner roles/contribution conflict'' that involves one partner ignoring or downplaying their partner's contributions. To accurately detect diverse types of conflict dialogues that emerge during collaborative learning, more contextual information is needed. Our experimental results showed that such information could be better captured by transformer-based models, such as RoBERTa, than by TF*IDF. 

In the audio modality, we found that pitch is the most predictive acoustic-prosodic feature for detecting Conflict. This result is aligned with prior research, which suggests that rapport building among learners is positively related to their pitch synchrony \citep{lubold2014acoustic}; prior research also suggests that when listeners express disagreement, their pitch levels tend to be higher than the speakers' pitch levels \citep{ma2022detecting}. In addition, all acoustic features yielded poor performance in detecting confusion. 

In the video modality, we found that facial AUs are the most predictive visual features for detecting Confusion compared to eye gaze and head pose; however, all visual features yielded poor performance for detecting Conflict. Specifically, we observed which facial AUs could indicate a speaker's expression of confusion. The top five detected facial AUs by the OpenFace analysis toolkit in our corpus for Confusion are AU1, AU4, AU23, AU12, and AU15. The top five detected facial AUs for Conflict are AU1, AU2, AU5, AU4, and AU23. Compared to these two categories, the top five detected facial AUs for Other are AU1, AU2, AU15, AU20, and AU25. This suggests that the distinctive facial AU patterns for detecting confusion were AU 4 (Brow Lowerer), AU 23 (Lip Tightener), and AU 12 (Lip Corner Puller). Taking together, it becomes evident that valuable cues for discerning states of confusion and conflict can be extracted from both audio and video modalities. However, a closer examination of the results reveals that prosodic information derived from the audio modality emerges as a stronger predictor of conflict, while the intricacies of facial Action Units (AUs) extracted from the video modality are more predictive of confusion (as detailed in Table \ref{tab:feature-selection}). These results showed the multifaceted nature of learners' collaborative dialogues, where nuances in both auditory and visual cues contribute uniquely to the understanding of their cognitive-affective states.

Our second research question asked: \textit{Does multimodal feature fusion help improve the detection performance compared to using data from a single modality? If so, what are the best multimodal combinations among those we considered?} We found that the best-performing model (F1 score of 0.59 for detecting confusion and 0.62 for detecting conflict) combined information from all three modalities. We also found that adding nonverbal features provided substantial performance improvement (F-1 scores increased by 0.16 and 0.19 for detecting confusion and conflict, respectively). We also investigated the potential influence of different multimodal feature fusion methods by comparing the performance among multimodal models using traditional Early and Late Fusion with another neural network-based cross-attention networks. These results indicated that neural-network-based multimodal fusion methods have a stronger capability for capturing the correlations and dependencies between modalities compared to non-neural-network-based fusion methods.

\subsection{Comparison with the State-of-the-Art}
Due to the lack of research on the automatic detection of confusion and conflict moments during collaboration, there is no available prior work to serve as a baseline for comparing the models reported in this paper. However, we can make three comparisons that contextualize the performance of these models. First, \citet{d2008automatic} investigated the automatic detection of individual learners' affective states during tutoring between a student and a spoken dialogue system. The model detected affective states (e.g., boredom, confusion, flow) using speech act features, and the maximum F-1 score in detecting confusion was 0.68. Second, \citet{grawemeyer2017affective} investigated the automatic detection of individual learners' affective states while engaging with an intelligent learning environment. The affect detection model had a precision score of 0.80 and a recall score of 0.11 for detecting confusion. Third, \citet{park2021detecting} developed automated models to detect talk that purposefully disrupted team dynamics and problem-solving interactions during collaborative game-based learning. The model used multimodal data, including learners' chat messages, gender, and pre-test scores. The maximum F-1 score in detecting disruptive talk was 0.70, with a precision score of 0.68 and a recall score of 0.74. Although the research context, goal, and modeling framework differ between their work and this study, our best-performing multimodal models trained with linguistic, acoustic, and visual features achieved comparable F-1 scores for detecting both confusion and conflict. 

To further understand the performance of the adaptive cross-attention fusion framework, Table~\ref{tab:confusion-matrix} shows the confusion matrix of the model.

\begin{table}[hbt!]
\caption[Confusion matrix of adaptive models.]{Confusion matrix of the model trained with adaptive cross-attention fusion framework. Among all misclassified cases (1,045 in total), the most dominant error is Other (actual) $\rightarrow$ Conflict (predicted), accounting for 51\% among all error cases.}
\centering
\label{tab:confusion-matrix}
\begin{tabular}{clccc}
\toprule
\multicolumn{1}{l}{} &  & \multicolumn{3}{c}{Predicted} \\ \hline
\multicolumn{1}{l}{} &  & \multicolumn{1}{l}{Confusion} & \multicolumn{1}{l}{Conflict} & \multicolumn{1}{l}{Other} \\
\multicolumn{1}{c}{\multirow{3}{*}{Actual}} & Confusion & 366 & 12 & 48 \\
\multicolumn{1}{c}{} & Conflict & 61 & 665 & 198 \\
\multicolumn{1}{c}{} & Other & 193 & 492 & 7867 \\ \bottomrule
\end{tabular}
\end{table}

As shown in the table, the model misclassified 1,045 out of 9,943 total samples. Within all 1,045 misclassified samples, the model performed very well in differentiating Confusion and Conflict, as indicated by 61 (5.8\%) error cases of Confusion (actual) $\rightarrow$ Conflict (predicted), and 12 (1.1\%) error cases of Conflict (actual) $\rightarrow$ Confusion (predicted). The model also performed well in differentiating Confusion and Other. This may be a result of the relatively simple structure of the confusion cases in our corpus, mainly including cases like: ``I don’t know,'' ``We need help,'' and ``I’m confused.'' Overall, the most dominant error type is Other (actual) $\rightarrow$ Conflict (predicted), accounting for 492 (47.08\%) among all error cases. This indicates that the model performed relatively poorly in differentiating Conflict and Other and was likely to predict an Other sample as Conflict. This result is also aligned with the main disagreement types among our human annotators, in which our two annotators mostly disagreed on Other and Conflict categories (See Table~\ref{tab:error-cases}), and we found that our automatic models tended to predict these controversial cases as Conflict instead of Other.

\begin{table}[hbt!]
\caption[Examples of controversial cases.]{Examples of controversial cases on which our human annotators disagreed in the corpus. Our multimodal models tended to predict these controversial cases as Conflict.} 
\centering
\label{tab:error-cases}
\begin{adjustbox}{width = 1\textwidth}
\begin{tabular}{@{}lccc@{}}
\toprule
\multicolumn{1}{c}{Example} & \begin{tabular}[c]{@{}c@{}}Annotator 1 \\ Grouped Category (Tag)\end{tabular} & \begin{tabular}[c]{@{}c@{}}Annotator 2\\  Grouped Category (Tag)\end{tabular} & Model Prediction \\ \midrule
Mm-mm, go to this one. & {\begin{tabular}[c]{@{}c@{}}Other \\ (Directive)\end{tabular}} & \begin{tabular}[c]{@{}c@{}}Conflict \\ (Disagreement/Negative Feedback)\end{tabular} & Other \\
It goes back up, so, that won't work. & Other & \begin{tabular}[c]{@{}c@{}}Conflict \\ (Disagreement/Negative Feedback)\end{tabular} & Conflict \\
No? & \begin{tabular}[c]{@{}c@{}}Other \\ (Question - Other)\end{tabular} & \begin{tabular}[c]{@{}c@{}}Conflict \\ (Disagreement/Negative Feedback)\end{tabular} & Conflict \\
I don't think- & \begin{tabular}[c]{@{}c@{}}Other \\ (Suggestion/Alternative Idea)\end{tabular} & \begin{tabular}[c]{@{}c@{}}Conflict \\ (Disagreement/Negative Feedback)\end{tabular} & Conflict \\
No we are working on it & \begin{tabular}[c]{@{}c@{}}Conflict \\ (Disagreement/Negative Feedback)\end{tabular} & Other & Conflict \\ \bottomrule
\end{tabular}
\end{adjustbox}
\end{table}

\subsection{Implications}
A main thrust of our work was to automatically detect confusion and conflict with supervised learning models and to identify the best multimodal feature combinations from among those we considered. This study provides a few important implications to advance research toward deploying intelligent learning systems for supporting collaborative learning.

First, our best-performing model combined features from lexical, audio, and video modalities, indicating that combining features from all three modalities improves the detection performance for confusion and conflict. However, experimental results comparing the performance among models trained with different unimodal models revealed that linguistic cues provided a strong performance foundation for the detection of both confusion and conflict, probably due to our manual annotation being text-based. In addition, adding non-verbal cues provided substantial performance improvement. Specifically, acoustic cues are more predictive of conflict, and facial cues are more predictive of confusion. This finding implies that for the future deployment of adaptive learning systems, combining information from the lexical and video modality may provide enough useful features to detect confusion by capturing predictive facial AUs. For detecting conflict, information from the lexical and audio modalities by capturing pitch variance may provide strong performance.

Second, we compared the performance of multimodal models with two traditional feature fusion methods and one neural network-based method. The results suggested that neural-network-based multimodal fusion methods have a stronger capability for capturing the correlations and dependencies between modalities compared to non-neural-network-based fusion methods.

Lastly, we found that models trained with neural-network-based fusion approaches had higher accuracy variance (SD = 0.18) during the cross-validation process compared to non-neural-network-based fusion approaches (SD = 0.08). This indicated that although a series of methods to avoid model overfitting were used, such as dropout layers and $L_2$ regularizers, models trained with neural-network-based fusion approaches are still more prone to  overfitting compared to non-neural-network-based fusion approaches due to more complicated model architectures. This problem can lead to poor generalizability, and this issue could be further exacerbated when the sample size is relatively small. Therefore, in the future development of intelligent systems, larger-sized datasets are needed to fully exploit the potential of the best performance from neural-network-based fusion approaches.v

\subsection{Limitations}
The current work has several important limitations. First, this study is limited to co-located pair-programming contexts in classroom environments and a relatively small corpus with 38 elementary school learners, and may not generalize to other learner groups. The predictive features found in this paper may not generalize well to group collaboration involving three or more team members or to learners in other populations or learning environments. Second, the naturalistic classroom setting limited the effectiveness of the video-derived features identified in this study. OpenFace relies heavily on a controlled video recording setup and sometimes fails to detect both learners' faces when they are not directly facing the camera or in the case of occlusion. Third, in this study, we did not consider the information of silences, which could provide valuable non-verbal information about the confusion of the last speaker. Last, some steps in the proposed framework are manual, such as the alignment the visual features after two learners switched seats during a learning session.

\section{Conclusion and Future Work}\label{sec9}
Confusion and conflict moments during collaborative learning may be beneficial for learning, yet may also generate frustration and impact the collaborative learning experience if they persist. Automatically detecting confusion and conflict during collaboration holds great promise for intelligent systems to provide learners with adaptive support. In this paper, we reported on automated models to detect confusion and conflict and compared a set of state-of-the-art multimodal learning analytics techniques with features extracted from learners' collaborative dialogues, including speech and facial behaviors. The experimental results revealed a series of important verbal and non-verbal indicators for detecting confusion and conflict; specifically, prosodic cues (pitch synchrony) are predictive of conflict, and facial cues (facial AUs) are predictive of confusion. In addition, the multimodal lexical + audio + video model with the early fusion strategy performed the best for this task.

There are several important directions for future work. First, future work should examine the generalizability of the findings in this study using larger datasets, including data from online learning environments and multi-party interactions among groups of three or more learners. To create a confusion and conflict detector that could be used in many different learning scenarios, it will also be important to determine how the predictive features of confusion and conflict differ with learners of different ages and cultures. Second, although OpenFace supports accurate detection of head pose, it remains challenging to integrate them into intelligent learning support systems for real-time analysis. Future work should investigate other methods and tools to detect learners' facial features accurately and time-efficiently. Third, it is important to compare difference feature fusion methods. Finally, more work is needed to investigate the effectiveness of feedback delivery strategies in adapting to learners' confusion and conflict moments during the collaborative learning process. Future work should also improve the data quality of audio and video collected in noisy classroom environments, such as using noise-canceling headsets and video techniques such as background blurring to help separate the intended targets from the background.

As we move toward detecting confusion and conflict in real time, we will be able to build and investigate systems that can significantly improve learners' collaborative learning experiences in classrooms. 

\section*{Acknowledgments}
This research was supported by the National Science Foundation through grants DRL-1721160. Any opinions, findings, conclusions, or recommendations expressed in this report are those of the authors, and do not necessarily represent the official views, opinions, or policy of the National Science Foundation.

\bibliographystyle{unsrtnat}
\bibliography{references}

\end{document}